\documentclass[aps,superscriptaddress,showpacs,twocolumn,
groupedaddress,nofootinbib,nobalancelastpage,nobibnotes]{revtex4}

\usepackage{amsmath}
\usepackage{amssymb}
\usepackage{epsfig}
\usepackage{graphicx}

\newcommand{\bi}{\begin{itemize}}
\newcommand{\ei}{\end{itemize}}

\newcommand{\be}{\begin{equation}}
\newcommand{\ee}{\end{equation}}
\newcommand{\bea}{\begin{eqnarray}}
\newcommand{\eea}{\end{eqnarray}}

\newcommand{\ldm}{\Delta m_{31}^2}
\newcommand{\sdm}{\Delta m_{21}^2}
\newcommand{\deltacp}{\delta_{\mathrm{CP}}}
\newcommand{\stheta}{\sin^2 2 \theta_{13}}

\newcommand{\ie}{{\it i.e.}}

\newcommand{\eg}{{\it e.g.}}

\newcommand{\cf}{{\it cf.}}
\newcommand{\etc}{{\it etc.}}
\newcommand{\eq}{Eq.}
\newcommand{\eqs}{Eqs.}

\newcommand{\fig}{Fig.}

\newcommand{\figs}{Figs.}

\newcommand{\Ref}{Ref.}
\newcommand{\Refs}{Refs.}
\newcommand{\Sec}{Sec.}
\newcommand{\Secs}{Secs.}

\newcommand{\Tab}{Table}

\newcommand{\JHFSK}{{\sc JHF-SK}}
\newcommand{\NUMI }{{\sc NuMI}}

\newcommand{\ReactorII}{{\sc Reactor-II}}
\newcommand{\JHFHK}{{\sc JHF-HK}}

\newcommand{\equ}[1]{\eq~(\ref{equ:#1})}
\newcommand{\figu}[1]{\fig~\ref{fig:#1}}

\begin{document}

\title{Understanding CP phase-dependent measurements at neutrino superbeams in terms of bi-rate graphs}

\author{Walter Winter}
\email[E-mail address: ]{wwinter@ph.tum.de}
\affiliation{Physik-Department, T30d,
Technische Universit{\"a}t M{\"u}nchen, James-Franck-Stra\ss{}e,
85748 Garching, Germany}

\date{\today}

\begin{abstract}
\vspace*{0.2cm}
We discuss the impact of the true value of the CP phase on the mass hierarchy, CP violation, and CP precision measurements at neutrino superbeams and related experiments. We we use a complete statistical experiment simulation including spectral information, systematics, correlations, and degeneracies to produce the results. However, since it is very complicated to understand the results in terms of a complete experiment simulation, we show the corresponding bi-rate graphs as useful tools to investigate the CP phase-dependencies qualitatively. Unlike bi-probability graphs, which are based upon oscillation probabilities, bi-rate graphs use the total event rates of two measurements simultaneously as a function of the CP phase. Since they allow error bars for direct quantitative estimates, they can be used for a direct comparison with a complete statistical experiment simulation.
We find that one can describe the CP phase dependencies of the mentioned measurements at neutrino superbeam setups, as well as one can understand the role of the $\mathrm{sgn} (\ldm)$-degeneracy. As one of the most interesting results, we discuss the dependence of the CP precision measurement as a function of the CP phase itself, which leads to ``CP patterns''. It turns out that this dependence is rather strong, which means that one has to be careful when one is comparing the CP precisions of different experiments. 
\end{abstract}


\maketitle

\section{Introduction}

 The leading atmospheric and solar neutrino oscillation parameters are now 
measured with unprecedentedly high precisions. The Super-Kamiokande~\cite{Fukuda:1998mi,Fukuda:1998ah,Toshito:2001dk}, SNO~\cite{Ahmad:2001an,Ahmad:2002jz,Ahmed:2003kj}, and KamLAND~\cite{Eguchi:2002dm}
experiments have especially contributed to these achievements. Furthermore, the establishment of the LMA (Large Mixing Angle) solution makes sure that three-flavor effects are not entirely suppressed by the mass hierarchy. This means that they could be detectable by the next-generation experiments -- provided that the leptonic mixing angle $\theta_{13}$ is large enough. So far, $\theta_{13}$ has been constrained to $\stheta \lesssim 0.1$ by the CHOOZ experiment~\cite{Apollonio:1999ae,Apollonio:2002gd}.  In order to determine $\theta_{13}$, the neutrino mass hierarchy, and the leptonic CP phase $\deltacp$, conventional beam experiments~\cite{Nakamura:2001tr,Paolone:2001am,Duchesneau:2002yq} are operated or built, and superbeam experiments~\cite{Itow:2001ee,Ayres:2002nm,Beavis:2002ye} are proposed. As an alternative, new reactor experiments with near and far detectors could provide some information on $\theta_{13}$ on a relatively short timescale~\cite{Martemyanov:2002td,Minakata:2002jv,Huber:2003pm,Shaevitz:2003ws}. After the superbeam era, future neutrino factories could find $\theta_{13}>0$ and test leptonic CP violation even below $\sin^2 2 \theta_{13} \sim 10^{-3}$ (for a summary, see \Ref~\cite{Apollonio:2002en}).

The next generation of experiments which are sensitive to {\em all} of the unknown oscillation parameters are the neutrino superbeams. In this study, we especially consider the JHF (J-PARC) to Super-Kamiokande (or Hyper-Kamiokande)~\cite{Itow:2001ee} and NuMI~\cite{Ayres:2002nm} off-axis superbeams, since specific LOIs or proposals exist for these experiments. Unfortunately, the potential to access $\theta_{13}$, the neutrino mass hierarchy, and the leptonic CP phase $\deltacp$ simultaneously implies that superbeams are facing a  high-dimensional and very complex parameter space. In particular, the measurements are spoilt by multi-parameter correlations~\cite{Huber:2002mx} and intrinsic degeneracies, which are the $(\deltacp,\theta_{13})$~\cite{Burguet-Castell:2001ez}, $\mathrm{sgn}(\Delta m_{31}^2)$~\cite{Minakata:2001qm}, and $(\theta_{23},\pi/2-\theta_{23})$~\cite{Fogli:1996pv} degeneracies leading to an overall ``eight-fold'' degeneracy~\cite{Barger:2001yr}. In principle, one could either combine two first-generation superbeams, or one superbeam with a reactor experiment to resolve these degeneracies for $\stheta \gtrsim 10^{-2}$~\cite{Wang:2001ys,Whisnant:2002fx,Barger:2002rr,Huber:2002rs,Minakata:2002jv,Minakata:2003ca,Migliozzi:2003pw,Huber:2003pm},
since these experiment types are proposed for comparable timescales and $\stheta$-values. If one wants to resolve the degeneracies with superbeams in the range $\stheta \sim 10^{-2} - 10^{-3}$, superbeam upgrades, such as the JHF (J-PARC) to Hyper-Kamiokande setup, will be required~\cite{Barger:2002xk,Burguet-Castell:2002qx,Donini:2002rm,Autiero:2003fu,Asratyan:2003dp}. For a systematic discussion about possibilities to resolve degeneracies on different scales of $\stheta$, see \Ref~\cite{Winter:2003st} and references therein.

The simulation techniques for future long-baseline experiments are very advanced now. We use in this study a complete statistical simulation including spectral information, a realistic detector simulation, systematics, correlations, and degeneracies. However, the results from such a simulation are not always trivial to understand, especially because of the impact of correlations and degeneracies. Therefore, it becomes more and more important to have tools to interpret these results. Note that these tools will never be able to replace the complete numerical simulations, but they can help to understand at least some of the aspects of the results. In this study, we use one of these tools, the bi-rate graphs, which are useful to understand the CP phase-dependencies of many measurements. The impact of $\deltacp$ is rather large in the considered experiments, since many of the problems with correlations and degeneracies originate in the correlation with $\deltacp$. We will mainly discuss the experiment performances as function of the ``true value of $\deltacp$''. We refer to the true value of $\deltacp$ as the simulated value,  which is realized by nature and which is, up to know, completely unknown.

Bi-rate graphs are based upon the so-called ``bi-probability'' graphs used in \Refs~\cite{Minakata:2001qm,Minakata:2001rj,Minakata:2002qe,Minakata:2002qi,Minakata:2003ca,%
Minakata:2003jj,Minakata:2003ma}.
In these graphs, the CP phase-dependence results in ellipses in the space of the oscillation probabilities of the two considered oscillation channels. Comparing the ellipses for different values of $\stheta$ and for different degenerate solutions leads to qualitative descriptions of many CP phase-dependent effects. In this work, we use the bi-rate space of the total charged-current event rates instead of the oscillation probabilities, such as it was first suggested in \Ref~\cite{Minakata:2001qm}. Since error bars can be used in this approach, semi-quantitative statements can be directly compared to a complete statistical analysis including systematics, multi-parameter correlations, and degeneracies. Because the off-axis superbeams are ``quasi-monochromatic'' beams due to their narrow beam spectrum (``narrow band beams''), spectral information plays only a minor role. Therefore, we will demonstrate that one can describe many properties of the results of complete statistical superbeam simulations  with the help of bi-rate graphs, and thus the underlying bi-probability picture. 

Besides the combination of superbeams, we will discuss the results including a new proposed reactor experiment~\cite{Huber:2003pm} in some cases, because we will observe that the results have a similar structure. We do not investigate future neutrino factories or other high-precision instruments, because the interpretation of the bi-rate approach is much more complicated for beams with good spectral information. Note that we treat the chosen experiments as examples in this study, which means that we do not want to give selection criteria for specific setups. The purpose of this study is to understand the CP phase-dependencies of complete statistical simulations in terms of bi-rate graphs phenomenologically. In fact, we will observe that the qualitative main results to not depend on the chosen setups.

This study is organized as follows: In \Sec~\ref{sec:biprob}, we analytically introduce the bi-probability picture and summarize some of its basic properties.  This section can be skipped by the reader familiar with the bi-probability approach. The complete statistical and bi-rate experiment simulations are described in \Sec~\ref{sec:sim}, where the experiment setups used as examples are introduced, too. As the main part of this work, in \Secs~\ref{sec:mass}, \ref{sec:cpviol}, and \ref{sec:cpprec} the CP phase-dependent properties of the mass hierarchy, CP violation, and CP precision measurements are discussed. Finally, we summarize our observations in \Sec~\ref{sec:summary}.

\section{General properties of bi-probability graphs}
\label{sec:biprob}

Bi-probability graphs, as they were introduced in \Refs~\cite{Minakata:2001qm,Minakata:2001rj,Minakata:2002qe,Minakata:2002qi,Minakata:2003ca,%
Minakata:2003jj,Minakata:2003ma},
show the probabilities of two measurements $P_1$ and $P_2$ simultaneously as a function of $\deltacp$. For a given set of parameter values, the oscillation probabilities $P_1$ and $P_2$ lead to a point in the bi-probability space $(P_1,P_2)$. Varying the CP phase from $0$ to $2 \pi$, one obtains ellipses with certain characteristics, which allow to study many properties depending on $\deltacp$. A very nice introduction to bi-probability graphs and their properties illustrated by simple figures can, for example, be found in \Ref~\cite{Minakata:2003jj}. Using total event rates instead of oscillation probabilities, one can also draw bi-rate graphs with similar properties. The advantage of bi-rate graphs is the possibility to use error bars in order to obtain direct quantitative estimates of many effects~\cite{Minakata:2001qm}. Many of the features introduced below can, for example, be found in \figu{anycpviolSKNUMI} (left), which may serve for illustration.
Typical cases for the measurements described by $P_1$ and $P_2$ are:
\begin{enumerate}
\item
 The probabilities $(P^{(1)}_{\mu e},P^{(2)}_{\mu e})$ of two different superbeam experiments, \ie, for two different values of $L/E$.
\item
 The CP-conjugated channels  $(P_{\mu e},P_{\bar{\mu} \bar{e}})$ of one superbeam experiment.
\item
The T-conjugated channels $(P_{\mu e},P_{e \mu})$ for the investigation of T violation.
\end{enumerate}
In this study, we will not consider case~3, since the fraction of $\nu_e$  produced by pion or kaon decays is rather small for superbeams.

In order to obtain some qualitative analytical understanding of bi-probability graphs, we can write the oscillation probabilities for a small $\theta_{13}$ and a small mass hierarchy parameter $\alpha \equiv \Delta m_{21}^2/|\Delta m_{31}^2|$ as~\cite{Minakata:2002qe,Minakata:2003ca}
\be
 P_i = X_i \, \sin^2 \theta_{13} + Y_i \, \sin \theta_{13} \, \cos \left( \deltacp \pm \frac{\Delta _{13}^i}{2} \right) + P^{\odot}_i, \label{equ:p}
\ee
$i \in \{1,2\}$, with
\bea
 X_i & \equiv & s_{23}^2 \, (\Delta_{13}^i)^2 \, \frac{\sin^2 \left( |\Delta_{13}^i \mp a_i L_i |/2 \right)}{\left(|\Delta_{13}^i \mp a_i L_i|/2 \right)^2}, \label{equ:x}\\
 Y_i & \equiv & \pm 2 \, c_{12} \, s_{12} \, c_{23} \, s_{23} \, \Delta_{12}^i \, \Delta_{13}^i \, \frac{\sin \left( a_i L_i / 2\right)}{a_i L_i/2} \nonumber \\
& & \times \frac{\sin \left( |\Delta_{13}^i \mp a_i L_i |/2\right)}{|\Delta_{13}^i \mp a_i L_i |/2}, \label{equ:y}\\
P^{\odot}_i & \equiv & \frac{1}{4} \, c_{23}^2 \, \sin^2 2 \theta_{12} \, (\Delta_{12}^i)^2 \, \frac{\sin \left( a_i L_i /2 \right)^2}{(a_i L_i/2)^2}, \label{equ:pdot}
\eea
and
\be
 \Delta_{ab}^i \equiv \frac{|\Delta m_{ab}^2| L_i}{2 E_i}, \quad a_i \equiv \sqrt{2} G_F N_e^i. \label{equ:delta}
\ee
In these equations, $i$ refers to the considered measurement $i \in \{1, 2\}$, $L_i$ to the baseline, $E_i$ to the neutrino energy, and $N_e^i$ to the electron number density in matter. In addition, $s_{ab} \equiv \sin \theta_{ab}$ and $c_{ab} \equiv \cos \theta_{ab}$. The upper signs in the individual terms correspond to neutrino running and a normal mass hierarchy.\footnote{Note that we only consider the oscillation channel $\nu_\mu \rightarrow \nu_e$ in this work.} For the inverted mass hierarchy, all the double signs change to the lower ones. For antineutrino running, the double signs in the $ |\Delta_{13}^i \mp a_i L_i |$-terms and in \equ{p} change to the lower ones. Note that using the inverted mass hierarchy as well as antineutrino running causes double sign changes at the respective places, \ie, some signs are changed back to the original ones.

In general, \equ{p} has the same structure as the original expansions in \Refs~\cite{Cervera:2000kp,Freund:2000ti,Freund:2001pn}: the $X_i$-term is proportional to $\sin^2 2 \theta_{13}$, the $Y_i$-term to $\sin 2 \theta_{13} \, \alpha$, and the $P_i^{\odot}$-term to $\alpha^2$. Thus, the CP effects are suppressed by the small $\theta_{13}$ and the mass hierarchy parameter $\alpha$. One can also proove that the CP trajectories are elliptic and that the minor and major axes of the ellipses are in case~2 in vacuum always parallel to the diagonal and its orthogonal for equal $L/E$ (\cf, Appendix of \Ref~\cite{Minakata:2001qm}). However, for different values of $L/E$, the pre-factors (offsets) in $P_1$ and $P_2$ can be different [\cf, \equ{x}] and the symmetry axis is not the diagonal anymore. A similar effect occurs for bi-rate graphs, because cross-sections, efficiencies \etc\ lead to different pre-factors even for the same $L/E$.

From \equ{p}, we can derive some basic properties of bi-probability graphs. For fixed oscillation parameters (except from $\deltacp$), this set of two equations gives the parameterized bi-probability trajectory in $(P_1,P_2)$-space as a function of $\deltacp$. It is obvious that the trajectory becomes degenerate if $P_1$ and $P_2$ have the same phases in the cosine term in \equ{p}. In this case, the ellipse simply collapses to a line. In order to create a phase shift between $P_1$ and $P_2$ in this term, one can, for example, use a different $L/E$ for $P_1$ and $P_2$ (different experiments, case~1), or neutrinos in $P_1$ and antineutrinos in $P_2$ (CP-conjugated channels, case~2). The special case of a phase shift $\pi/2$ results in a circle, such as it is obtained for one experiment in the oscillation maximum and the other in the oscillation minimum in case~1. The positions of the ellipses are determined by the offset proportional to $\sin^2 \theta_{13}$ (except from the small contribution from $P_i^{\odot}$), and the lengths of the axes of the ellipses are proportional to $\sin \theta_{13}$. Thus, it is usually quite illustrative to show many different ellipses for different values of $\sin \theta_{13}$, such as it is done in \figu{anycpviolSKNUMI}, left. In addition, the solutions for the normal and inverted mass hierarchies are normally drawn in each figure in order to discuss the effects of the $\mathrm{sgn}(\ldm)$-degeneracy~\cite{Minakata:2001qm}. Though the ellipses for the normal and inverted mass hierarchies are identical in vacuum, they move away from the diagonal (for equal $L/E$'s in $P_1$ and $P_2$) in opposite directions for an increasing matter potential of one experiment. This asymmetry comes from the different sizes of the $ |\Delta_{13}^i \mp a_i L_i |$-terms for the different mass hierarchies  and experiments,  which modify the offsets [\cf, \equ{x}]. Since one point in the bi-probability space corresponds to several sets $(\theta_{13}, \deltacp)$, \ie, different ellipses, it makes sense to show the area of all possible bi-probability trajectories generated by all possible values of $\theta_{13}$ for a specific mass hierarchy, which is often called a ``pencil'' (\cf, \figu{anycpviolSKNUMI}, left). Thus, a unique determination of the set $(\theta_{13}, \deltacp)$ is sometimes impossible because of the correlation between $\theta_{13}$ and $\deltacp$. In addition, if the two pencils for the positive and negative mass hierarchies overlap, then the situation becomes even more complicated because of the opposite-sign solutions.

In \equ{p}, the phase $\phi_i \equiv \Delta^i_{13}/2$ in the cosine term is $\pi/2$ at the first oscillation maximum.  Thus, close to the first oscillation maximum, we can make the approximation $\cos ( \deltacp \pm \phi_i ) \simeq \cos ( \deltacp \pm \pi/2) = \mp \sin \deltacp$. Assuming that $P_1$ and $P_2$ represent case~1 (two different experiments both using neutrinos), we can make some simple observations, which will be useful for the following sections:
\begin{description}
\item[Change of $\boldsymbol{L/E}$:] If $\phi_1=\phi_2$, \ie, for the same $L/E$, the ellipses collapse to lines. Moving from $\phi_1<\phi_2$ to $\phi_1>\phi_2$ (or vice versa) changes the orientation of the CP trajectories, as it can be easily seen from \equ{p}.
\item[CP phase closest to origin:] For $\phi_1 \sim \phi_2 \sim \pi/2$, \ie, both experiments operated close to the oscillation maximum,  $\cos ( \deltacp + \pi/2)$ is $- \sin \deltacp$ for the normal mass hierarchy. Thus, the CP trajectory is closest to the origin for $\deltacp \simeq \pi/2$ and farest away from the original for $\deltacp \simeq 3 \pi / 2$. For the inverted mass hierarchy, the additional sign change in $\phi_i$ is compensated by the sign change in $Y_i$ [\cf, \equ{y}], which means that we have the same behavior.
\item[Inverted mass hierarchy:] Since $\cos ( \deltacp - \phi) = \cos( - \deltacp + \phi)$ and the closest and farest values (from the origin) of the CP trajectory do not change for the inverted mass hierarchy, the CP trajectory changes its orientation. In addition, the $\sin \left( |\Delta_{13}^i \mp a_i L_i |/2 \right)/ \left( |\Delta_{13}^i \mp a_i L_i |/2 \right)$-terms in $X_i$ and $Y_i$ [\cf, \eqs~\ref{equ:x} and~\ref{equ:y}] become somewhat smaller for the inverted mass hierarchy close to the oscillation maximum, which means that the ellipses move closer to the origin and become somewhat smaller -- provided that matter effects are present. The same holds for the $P_i^{\odot}$-term in \equ{pdot}. Matter effects also cause the asymmetry between the positive- and negative-sign pencils coming from different $X_i$'s in \equ{x}.
\item[Effect of the solar $\boldsymbol{\sdm}$:] The size of the solar $\sdm$ changes the offsets [\cf, \equ{pdot}] as well as the axes lengths [\cf, \equ{y}] of the ellipses. Thus, the larger the value of $\sdm$ is, the more will the ellipses be extended.
\item[Very small values of $\boldsymbol{\theta_{13}}$:] For very small values of $\theta_{13}$, the smallness of $\theta_{13}$ dominates the splitting between the normal and inverted hierarchies, since in this limit the matter effect becomes a second order effect. Thus, the ellipses for the normal and inverted mass hierarchies strongly overlap.
\end{description}
Similar observations can be derived cases~2 and~3, which we however will 
not discuss further. However, it is interesting to note that for case~2 
the CP conserving solutions $0$ and $\pi$ lie on the line $P_1=P_2$ in vacuum because of the identical oscillation probabilities for neutrinos and antineutrinos. In addition, in this case usually an equal $L/E$ (same experimental setup) for $P_1$ and $P_2$ is assumed.

Since \equ{p} includes three terms, which can be of similar order of magnitude, other correlations than between $\deltacp$ and $\theta_{13}$ may be present and influence the measurement. So far, we have assumed that all parameters except from $\deltacp$ and $\theta_{13}$ are fixed. As long as the correlation between $\deltacp$ and $\theta_{13}$ dominates, the bi-probability picture will give the proper results. However, if there is a substantial contribution of another correlation, then there may be substantial differences to a complete statistical analysis including multi-parameter correlations. For example, the size of $\sdm$, and thus $P_i^{\odot}$, is obtained from solar neutrino experiments. Since it will not be exactly known at the time of the superbeam operation, the offset of the ellipses may vary within certain limits and lead to another error source. This example illustrates why we will find limitations of the bi-probability formalism in some cases. 

\section{Bi-rate picture and complete experiment simulation}
\label{sec:sim}

\begin{table*}[t!]
\begin{center}
\begin{tabular}{|l|c|c|c|}
\hline
& \JHFSK & \NUMI\ & \JHFHK \\
\hline
\multicolumn{4}{|l|}{Beam} \\
\hline
Target power& $0.77 \, \mathrm{MW}$ & $0.4 \, \mathrm{MW}$  & $4 \, \mathrm{MW}$ \\
$\nu$/$\bar{\nu}$-running & $\nu$ only & $\nu$ only & $\nu$ and $\bar{\nu}$ \\
Running period&5 years&5 years & 2 years ($\nu$)+6 years ($\bar{\nu}$) \\
\hline
\multicolumn{4}{|l|}{Detector} \\
\hline
Technology&Water Cherenkov&Low-Z calorimeter&Water Cherenkov\\
Fiducial mass&$22.5\,\mathrm{kt}$&$50\,\mathrm{kt}$&$1 \, 000 \,\mathrm{kt}$ \\
Baseline&$295\,\mathrm{km}$&$712\,\mathrm{km}$&$295\,\mathrm{km}$\\
Off-axis angle&$2^\circ$&$0.72^\circ$&$2^\circ$\\
\hline
\multicolumn{4}{|l|}{Overall facts} \\
\hline
Mean
$L/E$&$385\,\mathrm{km}\,\mathrm{GeV}^{-1}$&$320\,\mathrm{km}\,\mathrm{GeV}^{-1}$&$385\,\mathrm{km}\,\mathrm{GeV}^{-1}$\\
Mean energy&$0.76\,\mathrm{GeV}$&$2.22\,\mathrm{GeV}$&$0.76\,\mathrm{GeV}$\\
Signal events & $132.8$ & $334.4$ &  $12 \, 594$ ($\nu$) + $9 \, 239$ ($\bar{\nu}$) \\  
Background events & $22.8$ & $57.1$ & $2 \, 159$  ($\nu$) + $3 \, 352$ ($\bar{\nu}$) \\
\hline
\end{tabular}
\end{center}
\caption{\label{tab:base} The two first-generation superbeams \JHFSK\ and \NUMI , and the superbeam upgrade \JHFHK , as originally given in \Refs~{\rm \cite{Itow:2001ee,Ayres:2002nm}}, as well as the standard parameters used in this study. The total signal and background events are calculated for the appearance channels and the parameter values $\stheta=0.1$, $\deltacp=0$, $ |\ldm |=2.5 \cdot 10^{-3} \, \mathrm{eV}^2$, $\sin^2 2 \theta_{23}=1.0$, $\sdm=7.0 \cdot 10^{-5} \, \mathrm{eV}^2$,  $\sin^2 2 \theta_{12}=0.8$, and a normal mass hierarchy.}
\end{table*}

In this section, we first of all introduce the bi-rate approach as modification of the bi-probability picture, and discuss its limitations. Then, we introduce the complete statistical simulation techniques shortly, and point to earlier studies where they are described in greater detail. Eventually, we give the details of the experiment examples used in this study, as well as the oscillation parameters used.

In order to directly compare the bi-probability graphs to the results of a complete statistical analysis, we use the bi-rate picture. Instead of the bi-probability space, it shows the ellipses in the bi-rate space of the total (integrated) charged-current event rates.
 The main advantage of this approach is that statistical error bars can be used as a direct measure for the statistical significance of a specific result. Because of the off-axis technology to produce narrow-band beams,  the superbeams usually contain only very little spectral information. Thus, the total rates of the considered experiments carry the dominant physics information. Especially, the charged-current appearance rates dominate the measurements of the most interesting parameters $\stheta$, the mass hierarchy, and $\deltacp$.
Unfortunately, a small fraction of the detected electron neutrino (or antineutrino) signal is not originally part of the charged-current appearance rates and is often called ``background''. For the bi-rate approach, we do not use the background events, since they do not contain the interesting information (of course, we include them in the complete statistical simulation). 
Similar to the bi-probability graphs, one can show the total event rates in bi-rate space for any combination of two measurements. An example for a bi-rate graph can be found in \figu{anycpviolSKNUMI}, which we will use later. In this and similar figures, several ellipses for selected different values of $\stheta$ are shown, as it is described in the figure captions. The smaller $\stheta$ is, the smaller are the total rates and the closer are these ellipses to the origin. The different values of $\deltacp$ are marked by symbols, where the triangles refer to the CP conserving solutions. The total event rates for the normal mass hierarchy are shown with solid curves, where the regions of all possible ellipses (``positive-sign pencils'') are marked with the dark (green) shading.  The total rates for the inverted mass hierarchy are shown with dashed curves, where the regions of all possible ellipses (``negative-sign pencils'') are marked with the light (yellow) shading. At selected points, the statistical error bars for the total rates are shown at the 90\% confidence level. Note that they are only drawn for the directions along the axes in order to obtain an idea of their relative size. 

As we will see later, many properties depending on $\deltacp$ can be understood in terms of bi-rate graphs. However, this does not mean that a bi-rate graph can replace a complete experimental simulation. First of all, the experiment might be better than what one can see from the bi-rate picture, because there is at least some spectral information and in addition some information from the disappearance channels. For example, the $\mathrm{sgn}(\ldm)$-degenerate solution may actually not appear under the chosen confidence level. On the other hand, the experiment might be worse than the bi-rate interpretation, because the latter does not include systematics and other correlations than the one between $\stheta$ and $\deltacp$. However, we will be able to describe many CP phase-dependent properties of a complete simulation qualitatively, especially in the regions of the parameter space where the total appearance rates carry the dominant information. Therefore, the bi-rate picture can be used as a tool to understand many basic features of very complicated experiment simulations.

In order to demonstrate many effects quantitatively as realistic as possible, we use in parallel to the bi-rate approach a complete statistical analysis including systematics, multi-parameter correlations, and degeneracies. The corresponding analysis techniques are introduced and discussed in the Appendices of \Ref~\cite{Huber:2002mx} (general techniques) and \Ref~\cite{Huber:2002rs} (specifics of the superbeams), as well as in \Ref~\cite{Huber:2003pm} (reactor experiments). In order to use all available information, we fit the appearance and disappearance rates simultaneously, where we include the backgrounds as described in \Refs~\cite{Huber:2002mx,Huber:2002rs}. Therefore, we assume that the best measurement of the leading atmospheric parameters will be provided by the experiment (or experiment combination) itself at the time of the analysis. As external input, we assume the product $\sdm \cdot \sin 2 \theta_{12}$ to be measured with $15\%$ precision by the KamLAND experiment by the time the superbeam experiments will be analyzed~\cite{Barger:2000hy,Gonzalez-Garcia:2001zy}.

As examples for experiments, we study in this work the JHF (J-PARC) to Super-Kamiokande (or Hyper-Kamiokande)~\cite{Itow:2001ee} and the NuMI~\cite{Ayres:2002nm} projects, using the beams in the off-axis configurations. We refer to the first-generation superbeams with \JHFSK\ and \NUMI\ further on, as well as we use the JHF (J-PARC) to Hyper-Kamiokande superbeam upgrade \JHFHK . Their standard configurations, as they are proposed in the respective LOIs, are summarized in \Tab~\ref{tab:base}. Note that we use a \NUMI\ detector size of $50 \, \mathrm{kt}$ (as, for example, in the \NUMI\ proposal). We use five years of neutrino running time for each of the first-generation beams as well as two years of neutrino running and six years of antineutrino running for the \JHFHK\ upgrade in order to have approximately equal numbers of neutrino and antineutrino events. 
In addition, we use several modifications compared to \Tab~\ref{tab:base}: In order to be sensitive to the mass hierarchy, one needs a longer \NUMI\ baseline~\cite{Barger:2002xk,Huber:2002rs,Minakata:2003ca}. Therefore, we use a modified \NUMI\ baseline of $890 \, \mathrm{km}$ for the mass hierarchy measurement, which is the longest possible baseline for the fixed decay pipe and the considered off-axis angle. For the CP measurements, some CP-complementary information is needed to resolve the correlation between $\stheta$ and $\deltacp$. This can either be a substantial fraction of antineutrino running, or a large reactor experiment for a ``clean measurement'' of $\stheta$~\cite{Minakata:2002jv,Huber:2003pm,Minakata:2003ma,Minakata:2003wq}. Therefore, we either use \NUMI\ in the antineutrino running mode only, or the large reactor experiment \ReactorII\ from \Ref~\cite{Huber:2003pm} for CP measurements at the first-generation superbeams. However, note that for the first-generation superbeams the detection of CP violation at LMA-I will be very hard~\cite{Huber:2002rs}. Nevertheless, we will demonstrate that one could still learn something about the CP phase in some cases.

For the atmospheric oscillation parameters, we use, if not otherwise stated, $| \ldm |=2.5 \cdot 10^{-3} \, \mathrm{eV}^2$ and $\sin^2 2 \theta_{23}=1.0$~\cite{Toshito:2001dk}. For the solar oscillation parameters, we use $\sdm=7 \cdot 10^{-5} \, \mathrm{eV}^2$ and $\sin^2 2 \theta_{12}=0.8$ (see, for example, \Ref~\cite{Maltoni:2002aw}). Note that the LMA-II solution is disfavored by the SNO salt-enhanced neutral-current data~\cite{Ahmed:2003kj} (or, for example, \Ref~\cite{Maltoni:2003da}). For $\theta_{13}$, we only assume $\stheta \lesssim 0.1$~\cite{Apollonio:2002gd} below the CHOOZ bound and for $\deltacp$ we do not make and special assumptions, \ie,  $\deltacp \in [0, 2 \pi [$. In general, we include the full $(\deltacp, \theta_{13})$~\cite{Burguet-Castell:2001ez}, $\mathrm{sgn}(\Delta
m_{31}^2)$~\cite{Minakata:2001qm}, and $(\theta_{23},\pi/2-\theta_{23})$~\cite{Fogli:1996pv} degeneracies, which lead to an overall ``eight-fold'' degeneracy~\cite{Barger:2001yr}. However, since the energy resolution of the superbeams is rather poor and the atmospheric mixing angle is set to maximal mixing, the  $(\deltacp, \theta_{13})$- and  $(\theta_{23},\pi/2-\theta_{23})$-degeneracies are not relevant for this study.

\section{The determination of the mass hierarchy}
\label{sec:mass}

\begin{figure*}[t!]
\begin{center}
\includegraphics[width=16cm]{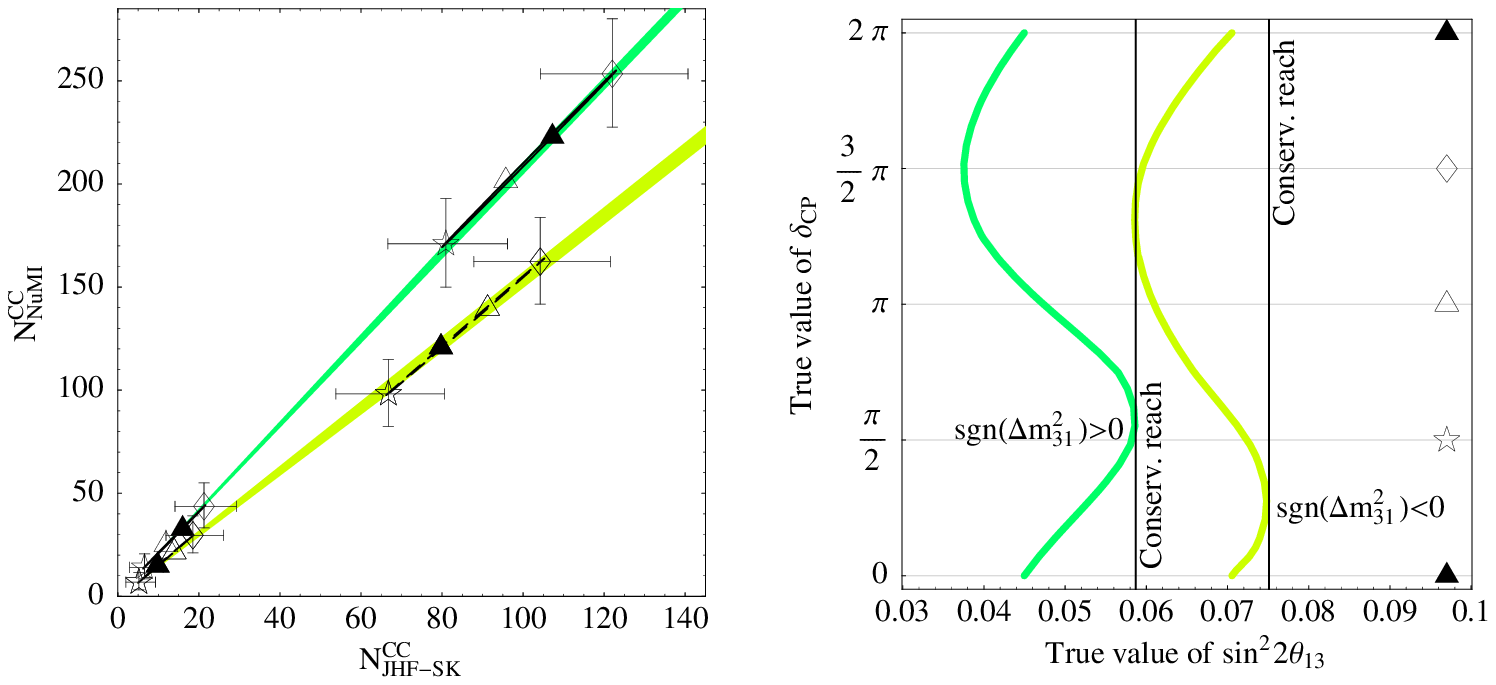}
\end{center}
\caption{\label{fig:sgncpdep} (Color online) The sensitivity to the mass hierarchy for \JHFSK\ and \NUMI @$890 \, \mathrm{km}$ combined, where both are operated with neutrino running only. The symbols on the ellipses correspond to $\deltacp=0$ (filled triangles), $\deltacp=\pi/2$ (stars), $\deltacp=\pi$ (unfilled triangles), and $\deltacp=3 \pi / 2$ (diamonds). The ellipses are shown for $\stheta=0.01$ and $0.08$, respectively, from the smallest to the largest, and the error bars for the total rates are drawn at the 90\% confidence level. The dark shaded (green) regions represent the positive-sign pencils with the solid curves, whereas the light shaded (yellow) regions represent the negative-sign pencils with the dashed curves. In the right-hand figure, the regions of sensitivity to a normal ($\mathrm{sgn}(\ldm )>0$) and inverted ($\mathrm{sgn}(\ldm )<0$) mass hierarchy are shown for the complete statistical experiment simulation including all information, where the sensitivity is given on the right-hand side of the curves at the 90\% confidence level. The horizontal lines are marked by the corresponding symbols from the left-hand plot, and the vertical lines correspond to the conservative reach in $\stheta$.}
\end{figure*}

We define sensitivity to a specific sign of $\ldm$ (\ie, the normal or inverted mass hierarchy) if there is no degenerate solution with the opposite sign of $\ldm$ fitting the simulated rate vector at the chosen confidence level. Since especially the $\mathrm{sgn}(\Delta m_{31}^2)$-degeneracy~\cite{Minakata:2001qm} implies that changing the sign of $\ldm$ can be often compensated by a different value of $\deltacp$, it constrains the potential of an experiment to determine the mass hierarchy. This already indicates that the true value of $\deltacp$ affects the mass hierarchy potential.  

We show in \figu{sgncpdep} the sensitivity to the normal and inverted mass hierarchy at the example of \JHFSK\ and \NUMI @$890 \, \mathrm{km}$ combined. In this figure, the bi-rate graph (left) is compared to the complete statistical simulation including spectral information, systematics, correlations, and degeneracies (right). Note that choosing a shorter \NUMI\ baseline would lead to a much weaker mass hierarchy sensitivity, as we have discussed in the last section. For our choice, the $L/E$'s of both experiments are very similar, which means that the bi-rate ellipses shrink to lines~\cite{Minakata:2003ca} and favor the mass hierarchy sensitivity. If the error bars at a chosen point in bi-rate (corresponding to a set of true parameter values) touch or overlap the opposite sign pencil, then the sensitivity to the positive sign of $\ldm$ will be destroyed at the chosen confidence level. Note that for a given $\stheta$, the true value of $\deltacp$  strongly influences the potential to establish the considered hierarchy, since the corresponding point may lie closer or farer away from the opposite-sign pencil. 
In addition, because of the correlation between $\deltacp$ and $\stheta$, any point within the opposite-sign pencil will destroy the sensitivity, which does not necessarily have the same values of $\deltacp$ and $\stheta$. In fact, one can identify the CP phase of the fake solution as the CP phase of the point closest to the original solution, since the total rate differences between those two points are smallest. This CP phase of the fake solution does usually not change as long as the pencils do not overlap. However, if they overlap, then it will start to move (\cf , \fig~8 of \Ref~\cite{Huber:2002mx}). 

One can obtain an estimate for the $\stheta$ sensitivity reach, \ie, the smallest value of $\stheta$ for which the considered sign of $\ldm$ can be established, by finding the ellipse for which the error bars just touch the opposite-sign pencil (for a fixed value of $\deltacp$). In \figu{sgncpdep}, this means that one obtains the sensitivity reach by shifting the ellipse down on the pencil until the error bars of a fixed point (fixed $\deltacp$!) touch the opposite-sign pencil. The sensitivity reach can then be read off as the corresponding value of $\stheta$ for which this ellipse is drawn. 
 Of course, this sensitivity reach depends on the true value of $\deltacp$, which determines the position on the ellipse. One can easily read off the bi-rate graph, that the best sensitivity reach is obtained close to $\deltacp=3 \pi/2$ and the worst close to $\deltacp=\pi / 2$.
This behavior can also be found in the complete simulation result -- it 
originates in the strong dependence of the total event rates on the true
value of $\deltacp$. 
In addition, the estimated sensitivities from the error bars in the bi-rate picture are a good estimate for the actual sensitivity reaches.
For the negative-sign pencil, the ellipses are (for the same values of $\stheta$) somewhat closer to the origin, which means that the sensitivity to the inverted mass hierarchy is somewhat worse. Nevertheless, the qualitative behavior as a function of the true value of $\deltacp$ can be rediscovered. 

There are two more interesting aspects in \figu{sgncpdep}: First, since we do not know the true value of $\deltacp$ realized by Nature, it is often convenient to show the conservative sensitivity reach in more condensed plots (\cf, vertical lines in right panel). This conservative reach then describes what region the experiment will be able to cover for sure, and helps to optimize it in terms of the other parameter values (such as in \figs~4 and~8 of \Ref~\cite{Huber:2002rs}). Second, as we have discussed in \Sec~\ref{sec:biprob}, the size of $\sdm$ is proportional to the sizes of the ellipses. Therefore, for a larger (smaller) $\sdm$ within the solar-allowed region, the ellipses become more (less) extended and the dependence on the true value of $\deltacp$ becomes stronger (smaller). Therefore, it can be shown that the amplitudes of the $\deltacp$-dependencies in \figu{sgncpdep}, right panel, depend on $\sdm$. However, mainly the conservative reaches (close to $3 \pi/2$) are affected for large $\sdm$, whereas the best-case reaches (close to $\pi/2$) remain almost unaffected.

\section{The sensitivity to CP violation}
\label{sec:cpviol}

\begin{figure*}[t!]
\begin{center}
\includegraphics[width=16cm]{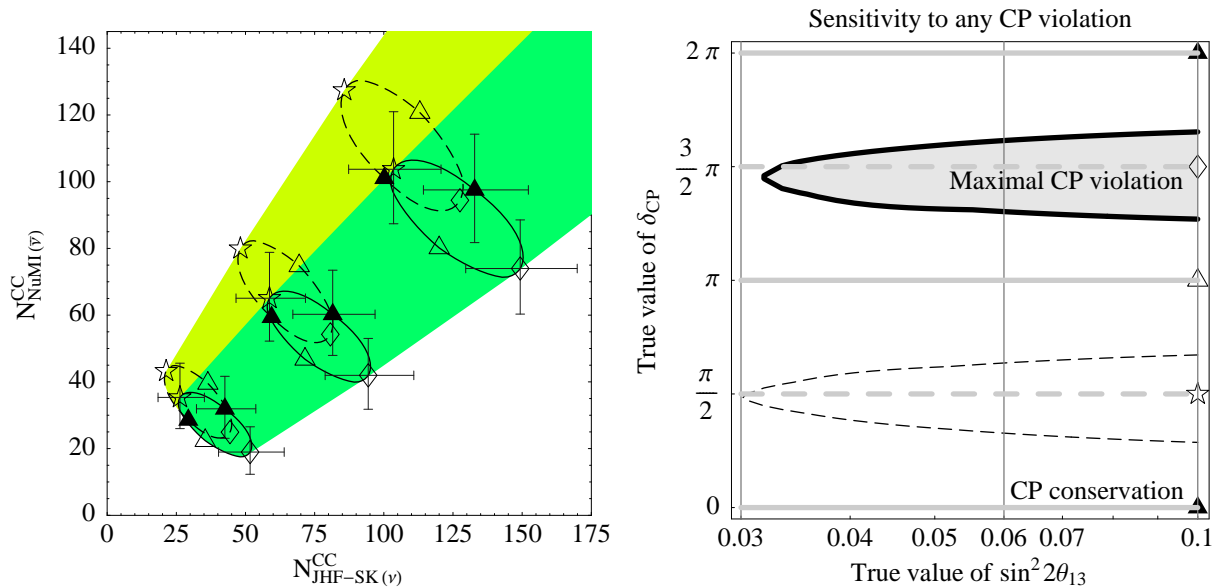}
\end{center}
\caption{\label{fig:anycpviolSKNUMI} (Color online) The sensitivity to {\em any}  CP violation ($\deltacp \neq$ $0$, $\pi$) for \JHFSK\ (neutrino running only) and \NUMI\ (antineutrino running only). The left-hand plot shows the bi-rate graph, where the ellipses are drawn for $\stheta=0.03$, $0.06$, and $0.1$, respectively, from the smallest to the largest. The right-hand plot marks the regions of sensitivity to CP violation as a function of the true values of $\stheta $ and $\deltacp $ for the complete statistical experiment simulation including all information, where the sensitivity is given within the gray-shaded region. The dashed curve refers to the additional sensitivity if one is not taking into account the degenerate solution. The vertical lines correspond to the values of $\stheta $ for which the ellipses in the left-hand plot are drawn. The horizontal lines are marked by the corresponding symbols from the left-hand plot, where solid lines refer to CP conservation and shaded lines to maximal CP violation. The symbols and other parameters are the same as in \figu{sgncpdep}, but we assume a normal mass hierarchy in the right-hand plot. For the inverted mass hierarchy, the role of the two peaks in the right-hand plot would be exchanged.}
\end{figure*}

We define sensitivity to CP violation if the experiment can distinguish $\deltacp$ from the CP conserving values $0$ and $\pi$ at the chosen confidence level. This means that for a fixed simulated/true value of
$\deltacp \notin \{ 0, \pi \}$, the CP violation sensitivity will be
 destroyed if any of the CP conserving values $0$ or $\pi$ fits the 
chosen simulated value $\deltacp$. Since, in principle, any value $\deltacp 
\notin \{ 0, \pi \}$ violates CP, one can discuss the CP violation
sensitivity as a function of this simulated value. 
Obviously, the closer
$\deltacp$ is to a CP conserving value $0$ or $\pi$, the harder it will 
be to establish the (in this case small) CP violation.
A special case of the sensitivity to CP violation is the sensitivity to {\em maximal} CP violation, which corresponds to the true values $\deltacp=\pi/2$ or $3 \pi / 2$. Since these values are farest away from CP conservation,
the CP violation will in most cases be easiest to detect. In addition, we 
will refer  to the CP violation for any simulated value $\deltacp \neq 0$, $\pi$ as the sensitivity to {\em any} CP violation.

Because of the correlation between $\stheta$ and $\deltacp$, it is necessary to have some information on the CP conjugated channels in order to efficiently measure $\deltacp$. It has, for example, been demonstrated in \Refs~\cite{Barger:2002xk,Huber:2002rs,Minakata:2003ca} that this implies a substantial fraction of antineutrino running in one or both of the \JHFSK\ and \NUMI\ off-axis experiments. Alternatively, a large reactor experiment could supply the CP-complementary information~\cite{Minakata:2002jv,Huber:2003pm,Minakata:2003ma,Minakata:2003wq}. However, it has been shown that CP violation measurements are very difficult for first-generation superbeams at the LMA-I solution. As we will demonstrate later, one of the best options for the superbeams is to operate \JHFSK\ with neutrinos and \NUMI\ with antineutrinos only. Therefore, we use this configuration in this section as an example, and we will discuss other possibilities in the next section. In fact, we will observe that none of the other alternatives can do a CP violation measurements.

We show in \figu{anycpviolSKNUMI} the potential to establish {\em any} CP violation for \JHFSK\ (neutrino running mode) and \NUMI\ (antineutrino running mode), \ie, the potential to establish
that a chosen value $\deltacp$ can be distiguished from the CP conserving solutions $0$ and $\pi$. In the left-hand panel, the bi-rate graph is shown
for the total charged current event rates of the two experiments, where the smaller antineutrino cross-section is compensated by the large \NUMI\ detector in order to have similar error bars with similar statistical weights. In the right-hand panel, the region of sensitivity to CP violation (within thick black curve) is given as a function of the true values of $\stheta$ and $\deltacp$ for the complete statistical experiment simulation including all information. One can read off the sensitivity to maximal CP violation $\deltacp=\pi/2$ or $3 \pi / 2$ along the horizontal dashed lines in the right-hand plot. 

Let us first of all discuss what one could learn about the sensitivity to {\em maximal} CP violation from the  bi-rate graph in \figu{anycpviolSKNUMI}, where we include the effect of the $\mathrm{sgn} (\ldm)$-degeneracy. For example, if we assume a normal mass hierarchy and $\deltacp=+\pi/2$ (stars), we will only be able to discover CP violation (at the chosen confidence level) if the error bars do not overlap the CP conserving values $0$ and $\pi$ (triangles). Thus, the CP violation sensitivity is characterized by making the CP conserving values $0$ and $\pi$ ``special'' and it is determined by the relative positions of the CP violating and CP conserving values considered. It is often not directly representative for a CP precision measurement which does not emphasize any special values of $\deltacp$, and vice versa. For example, a CP precision of $180^\circ$ at $\deltacp=\pi/2$ does not necessarily mean that one can establish maximal CP violation, since the fit manifold may not be exactly symmetric around $\deltacp=\pi/2$ and one of the CP conserving solutions may be overlapped.

The definition of the CP violation sensitivity also implies that overlapping one of the CP conserving values of any degenerate solution (which is present at the chosen confidence level) would destroy this sensitivity. For the  $\mathrm{sgn} (\ldm)$-degeneracy and a normal mass hierarchy, this corresponds to an overlap with the triangles of the dashed curves in the bi-rate graph. In our example, the negative-sign ellipses are moved towards $\deltacp=\pi/2$ and always from $\deltacp=3 \pi / 2$ on the positive-sign ellipses. This means that for  $\deltacp=\pi/2$, the degenerate solution determines the CP violation sensitivity and for $\deltacp=3 \pi / 2$, the original solution (for the normal mass hierarchy). Thus, the degenerate solution is only important for the sensitivity to $\deltacp=\pi/2$. In addition, this interplay between the original and degenerate solutions can easily be read off the bi-rate graphs for the inverted mass hierarchy and other setups. For example, for the inverted mass hierarchy, the role of  $\deltacp=\pi/2$ and  $\deltacp=3 \pi/2$ is just exchanged. However, it should be noted that for high-luminosity upgrades, the degenerate solution does not necessarily appear at all below the chosen confidence level. For the first-generation superbeams, it is normally present at very low $\chi^2$-values. Eventually, one can immediately read off the bi-rate graph, that the sensitivity to maximal CP violation is given for large values for $\stheta$ close to $\deltacp=3 \pi /2$ only, which is consistent with the complete analysis in the right-hand panel of \figu{anycpviolSKNUMI} (along horizontal dashed lines).  

One can also estimate the sensitivity to {\em any} CP violation $\deltacp\neq 0$, $\pi$ by following the points on a specific  ellipse in the bi-rate picture until the error bars touch one of the CP conserving solutions (triangles). With this approach, one can read off how close one can get to the CP conserving solutions. Indeed, the estimates from the bi-rate picture are consistent with the result from the complete experiment simulation in the right-hand panel of \figu{anycpviolSKNUMI}. One can easily see by the comparison of the dashed and thick solid curves in the right-hand panel that the impact of the $\mathrm{sgn}( \ldm )$-degeneracy is very large, since the strong matter effects at the \NUMI\ beam cause a large separation of the positive- and negative-sign pencils. 
It completely destroys the sensitivity to maximal CP violation $\deltacp=\pi/2$, while the sensitivity to maximal CP violation $3 \pi / 2$ is still given down to $\stheta \simeq 0.03$. In addition, the sensitivity to {\em any} CP violation is only present very close to the maximally CP violating value. Again, one can read off this figure that the $\mathrm{sgn}(\ldm )$-degeneracy only affects the first and second quadrants (for the positive mass hierarchy)~\cite{Minakata:2001qm}, since the negative-sign ellipses are shifted into the direction of $\deltacp=\pi/2$ for the considered setup. Thus, the CP conserving values are shifted towards $\deltacp=\pi/2$ and make the sensitivity to CP violation worse for any value within these quadrants. From the dashed curve in the right-hand figure without $\mathrm{sgn}(\ldm )$-degeneracy, it is obvious that the sensitivity to {\em any} CP violation is symmetric before the inclusion of this degeneracy. Thus, in this case, the $\mathrm{sgn}(\ldm )$-degeneracy is the main reason for the asymmetry of the CP sensitivities between the first/second and third/fourth quadrants.

The concepts in \figu{anycpviolSKNUMI} can also be extended to superbeam upgrades, such as \JHFHK . In fact, one can also observe that only the first and second quadrants (around $\deltacp = \pi/2$) are affected by the $\mathrm{sgn}( \ldm )$-degeneracy in a similar way to this figure. For \JHFHK , corresponding figures can be found as \fig~18 in \Ref~\cite{Huber:2002mx} or \fig~7 in \Ref~\cite{Ohlsson:2003ip}. Note that for neutrino factories, the behavior is somewhat different and more complicated, because they use a much broader energy spectrum and a different $L/E$.

\section{The precision of $\boldsymbol{\deltacp}$: ``$\boldsymbol{\deltacp}$-patterns''}
\label{sec:cpprec}

We define the precision of $\deltacp$ as the fraction of the CP circle which fits the selected true value of $\deltacp$ at the chosen confidence level. Therefore, we also call it ``CP coverage'', since it describes what fraction of the fit values is ``covered'' on the CP circle by an individual measurement, and allows to include complicated topologies in a very simple way. A coverage (or precision) of $360^\circ$ corresponds to all values of $\deltacp$ fitting the true value (and thus, no information on $\deltacp$) and a coverage (or precision) close to $0^\circ$ to a very precise measurement. Hence, the value of $360^\circ$ minus the CP coverage corresponds to the range which can be excluded. Any value of the precision which is smaller than $360^\circ$ will therefore imply some information on $\deltacp$ and exclude certain regions on the CP circle. From this definition, the difference to the CP violation sensitivities is obvious: The CP precision measurement answers the question if we can learn something about $\deltacp$ at all, and the CP violation measurement if we can distinguish CP violation from CP conservation. For example, for the case of $\deltacp$ close to $0$ or $\pi$, there is no sensitivity to CP violation, but we may still be able to exclude certain regions on the CP circle. For the CP precision measurement, the treatment of the $\mathrm{sgn}(\ldm)$-degeneracy is straightforward and is given by its definition: Any solution of the opposite-sign circle fitting the true value of $\deltacp$ at the chosen confidence level has to be removed from the allowed regions on the CP circle.

\begin{figure*}[t!]
\begin{center}
\includegraphics[width=15cm]{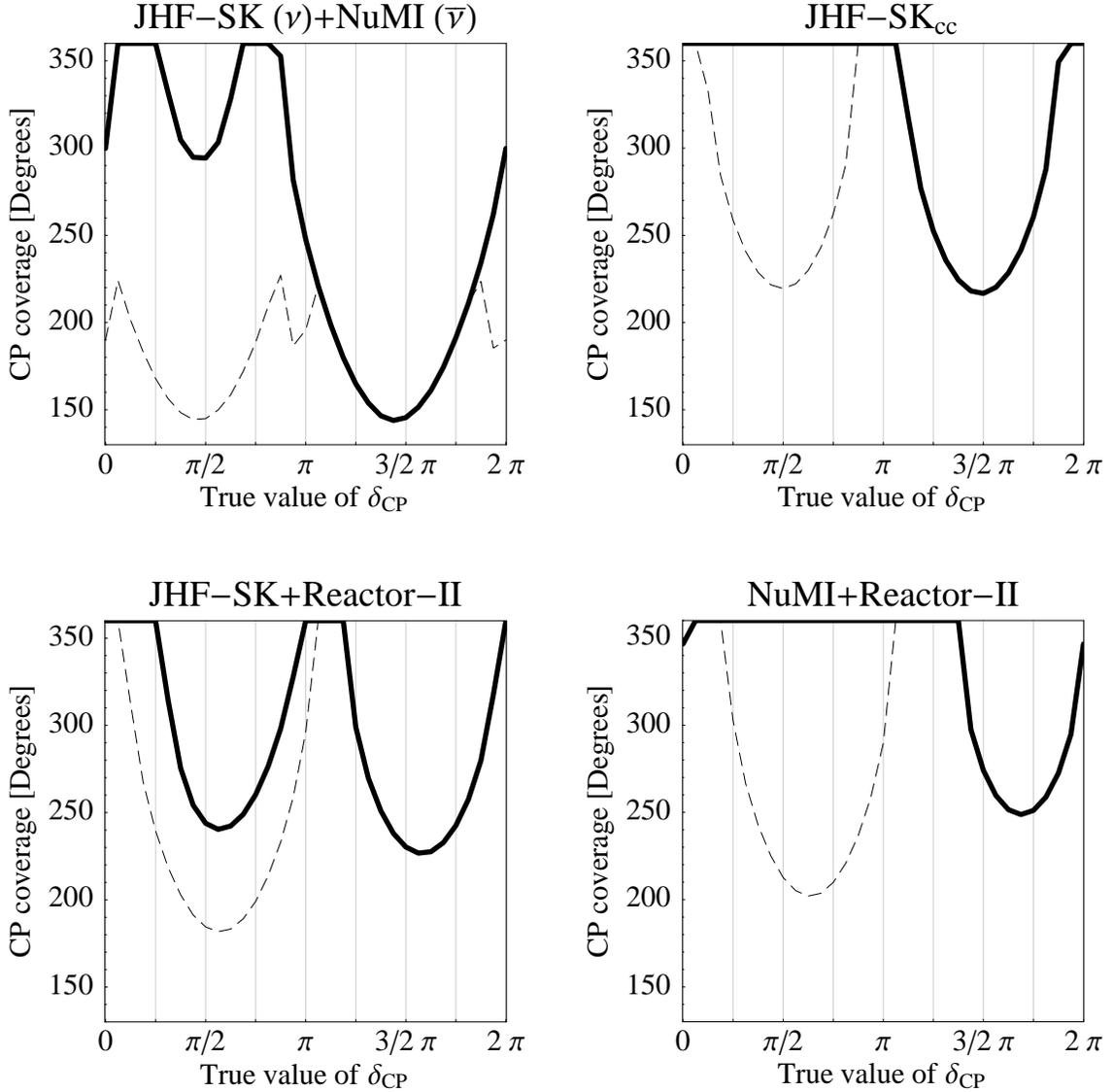}
\end{center}
\caption{\label{fig:cpdiffexp} The CP coverage as a function of the true value of $\deltacp$ for the four different experiment combinations as given in the plot labels and a complete statistical experiment simulation including all information (90\% confidence level): \JHFSK\ with neutrino running plus \NUMI\ with antineutrino running,  \JHFSK $_{\mathrm{cc}}$ in the combined neutrino-antineutrino running mode (25\% of the total running time in the neutrino mode and 75\% in the antineutrino running mode), \JHFSK\ plus \ReactorII , and \NUMI\ plus \ReactorII . Here \ReactorII\ refers to a large reactor experiment with near and far detectors, as it is defined in \Ref~\cite{Huber:2003pm} with an integrated luminosity of $8 \, 000 \, \mathrm{t} \, \mathrm{GW} \, \mathrm{y}$. In these plots, the CP coverage [Degrees] is defined as the part of the CP circle which fits the true value which is given on the horizontal axis, \ie, $360^\circ$ corresponds to no information on $\deltacp$.  For the oscillation parameters the current best-fit values with $\stheta=0.1$ and a normal mass hierarchy are assumed. The dashed curves refer to not taking into account degeneracies.}
\end{figure*}

In \figu{cpdiffexp}, we show the CP coverage as a function of the true value of $\deltacp$ for a complete statistical analysis including all information and different combinations of next-generation experiments as examples. Obviously, the precision strongly depends on the true value of $\deltacp$. Especially the dependence on the true value of $\deltacp$ allows to quote a wide range of values for the CP precision of a specific experiment or combination of experiments. One can also see that this dependence is very complicated, although there are some regularities. We further on refer to these regularities as a function of the true value of $\deltacp$ as ``CP patterns''. 

Let us first compare the CP coverage measurement in \figu{cpdiffexp} to the CP violation measurement. In order to have sensitivity to maximal CP violation, it is necessary to have a CP precision smaller than $180^\circ$ at the CP violating values $\deltacp=\pi/2$ or $3 \pi / 2$.\footnote{It is necessary, but not sufficient, since the fit manifold may not be symmetric around the maximally CP violating value and still overlap one of the CP conserving solutions.} Thus, only the first experiment combination (\JHFSK\ with neutrinos and \NUMI\ with the large detector and antineutrinos) could have some sensitivity to maximal CP violation (indeed, it has, as it can be seen from \figu{anycpviolSKNUMI}). All the other combinations do not have a sensitivity to CP violation, as it has been demonstrated in \Ref~\cite{Huber:2002rs} for combinations of first-generation superbeams. However, \figu{cpdiffexp} shows that one may still learn something about the CP phase if $\stheta$ is large, which strongly depends on the true value of $\deltacp$. In order to exclude certain ranges for $\deltacp$, the CP coverage has to be smaller than $360^\circ$. Since we do not know yet what the true value of $\deltacp$ is, one can be lucky or unlucky if one is building an experiment with large ranges with a CP coverage of $360^\circ$. Thus, the risk for such an experiment not to supply any new information is rather high.

In the following, we will give a qualitative description of the CP patterns of the next-generation experiments by comparing \figu{cpdiffexp}, upper left panel, with \figu{anycpviolSKNUMI}, left, since these two figures are computed for the same experiments and parameter values. Note that the largest ellipses in \figu{anycpviolSKNUMI} computed for $\stheta=0.1$ correspond to \figu{cpdiffexp}.
The other experiments in \figu{cpdiffexp} show the same qualitative behavior.\footnote{Especially, similar to the CP-conjugated channel at a superbeam, the reactor experiment can also help to resolve the correlation between $\stheta$ and $\deltacp$ with a precise $\stheta$ determination~\cite{Huber:2003pm}, which means that one would not expect large surprises in this case.} In order to keep this description simple, it is first of all useful to ignore the effect of the $\mathrm{sgn}(\ldm)$-degeneracy, \ie, to concentrate on the dashed curves. 
In \figu{anycpviolSKNUMI}, left, the error bars are relatively large compared to the size of the ellipses. In addition, systematics and correlations make them even larger than they appear there. Thus, for $\deltacp=0$ or $\pi$ the error bars cover the whole ellipse and one cannot obtain any information on $\deltacp$, \ie, the error in \figu{anycpviolSKNUMI} (left) reaches $360^\circ$. In particular, they always touch the opposite side of the ellipse, because the correlation with $\stheta$ allows an ellipse for a different value of $\stheta$ to overlap the original one. However, for $\deltacp=\pi/2$ or $3 \pi / 2$ only half of the error bars overlap the ellipse, which means that there the information on $\deltacp$ is significantly better. Depending on the size of the error bars, and thus the true value of $\stheta$, the CP coverage can differ up to a factor of two between $\deltacp=0$ or $\pi$ and $\deltacp=\pi/2$ or $3 \pi / 2$. 

As far as the $\mathrm{sgn}(\ldm)$-degeneracy is concerned, we illustrate its effects for a normal mass hierarchy by the dashed (no degeneracy) and solid (final result) curves in \figu{cpdiffexp}. Since, in this example, the error bars are rather large, it is only effective for the first and second quadrants in a straightforward way, because there ``new'' values of $\deltacp$ enter the error bars by the shifting of the negative mass hierarchy ellipse (\cf, \figu{anycpviolSKNUMI}, left). Because the positive and negative mass hierarchy ellipses are rather different from each other for large values of $\stheta$, but quite similar to each other for small values of $\stheta$, the effects of the degeneracy are much larger for the large values of $\stheta$ (provided that it is present at all at the chosen confidence level). For the inverted mass hierarchy, one would just observe the inverted behavior for the exchanged quadrants. Thus, the $\mathrm{sgn}(\ldm)$-degeneracy breaks the symmetry between $\deltacp=\pi/2$ and $3 \pi / 2$~\cite{Minakata:2001qm}.

\begin{figure*}[t!]
\begin{center}
\includegraphics[width=16cm]{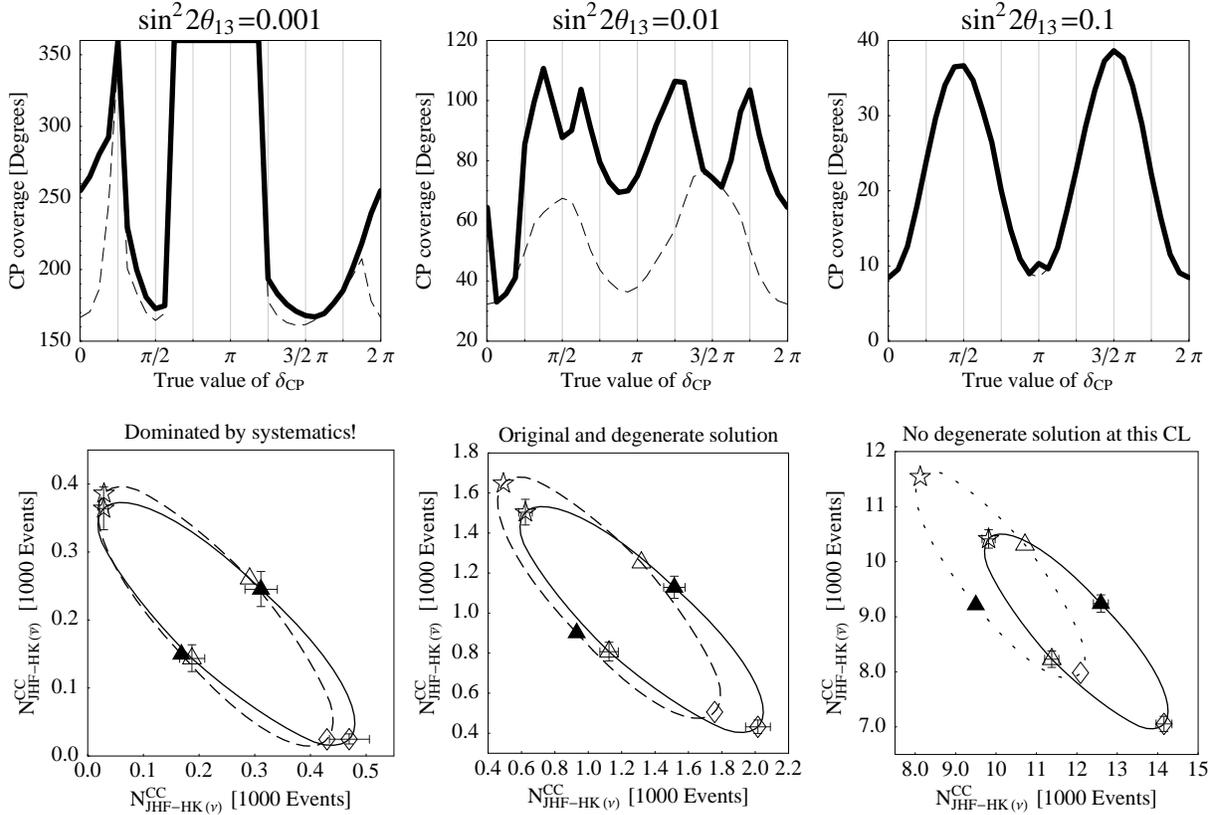}
\end{center}
\caption{\label{fig:hkcpprec} The CP coverage (upper row) as a function of the true value of $\deltacp$ for \JHFHK\ and three different values of $\stheta$ as given in the plot labels (90\% confidence level). The plots in the upper row include the complete statistical simulation including all information and correspond to the ellipses in the lower row. For the oscillation parameters the current best-fit values and a normal mass hierarchy are assumed. The dashed curves refer to not taking into account degeneracies. The meaning of the symbols is explained in \figu{sgncpdep}.}
\end{figure*}

Compared to exclusion measurements of the leptonic CP phase, it is interesting to investigate high-precision measurements, such as from a superbeam upgrade. We here focus on \JHFHK , which has a very narrow energy spectrum and is therefore well-suited for the bi-rate approach. In addition, the CP coverage is for very precise measurements identical to the precision and can therefore describe both exclusion and high-precision measurements. \figu{hkcpprec} shows the CP precision as a function of $\deltacp$ (upper row) for three different values of $\stheta$ (in columns). In the lower row of \figu{hkcpprec}, the corresponding ellipses are drawn. Note that \JHFHK\ is operated in the combined neutrino-antineutrino running mode with two years of neutrino running and six years of antineutrino running.

For \JHFHK , the statistical error bars are very small compared to the ellipses. Thus, without taking into account degeneracies (dashed curves in upper row), the CP patterns are, in principle, very regular (though different from the case of large error bars - see below). However, there are some complications. First, for $\stheta=0.001$, the superbeam upgrade is dominated by systematics, which means that the actual error bars (including both statistics and systematics) are much larger than the statistical error bars in the figure.  Second, the effects of the degenerate solution can either be rather large and may lead to very irregular patterns (\eg , for $\stheta=0.01$) or the degeneracy may not be present at all at the chosen confidence level (\eg , for $\stheta=0.1$). Comparing the ellipses to the corresponding CP pattern, we can distinguish three cases with respect to the size of the error bars and the $\mathrm{sgn}(\ldm)$-degeneracy:
\begin{description}
\item[Very small values $\boldsymbol{\stheta}$.] For small values of $\stheta$ (left-hand panel of \figu{hkcpprec}), the actual error bars are much larger that the statistical error bars in \figu{hkcpprec} (left) because of the systematics domination of the background and its uncertainty, as well as other effects.  Thus, we have a similar case to the first-generation superbeams, where the CP coverage is best close to $\pi/2$ and $3/2 \pi$ (without degeneracy). In addition, the $\mathrm{sgn}(\ldm)$-degeneracy is present at the chosen confidence level. However, the ellipses for the original and degenerate solutions are very similar to each other and the actual error bars are rather large, which means that the effect of the degeneracy is small. Note that this specific case is an example of the limitations of the bi-rate approach, since the systematics is more complicated to be simply included in the bi-rate picture.
\item[Large values of $\boldsymbol{\stheta}$.] For large values of $\stheta$ (see right column of \figu{hkcpprec}), \ie , $\stheta \gtrsim 0.01$, we obtain the
simplest pattern with a continuous dependence on $\deltacp$. Since in this case the error bars are very small compared to the size of the ellipses, the CP precision becomes best at $0$ and $\pi$ and worst at $\pi/2$ and $3/2 \pi$ because of the shallower dependence on $\deltacp$ close to $0$ and $\pi$. In addition, the $\mathrm{sgn}(\ldm)$-degeneracy is normally not present at the chosen confidence level. Thus, it hardly affects the CP pattern.
\item[Intermediate values of $\boldsymbol{\stheta}$.] For intermediate values of $\stheta$ (see middle column of \figu{hkcpprec}), the pattern without degeneracy correspond to the one for large $\stheta$ and is very regular (\cf, dashed curve). However, the degenerate solution is present at the chosen confidence level and can have a very large effect especially for $0.001 \ll \stheta \lesssim 0.01$ (\cf , \fig~8 of \Ref~\cite{Huber:2002mx}). Although its effect is, in principle, stronger in the first and second quadrants around $\deltacp=\pi/2$ than in the third and fourth quadrants (such as for \JHFSK ), it can also affect the other quadrants because the ellipses for the original and degenerate solutions are still very close to each other in this $\stheta$-range. Since the error bars are very small compared to the size of the ellipse, the CP coverage can be easily doubled 
if the degenerate solution moves ``new'' values of $\deltacp$ into the error bars of the original solution, \ie, the CP precision becomes worse. However,
in some cases (shortly after $0$ and close to $3 \pi / 2$) the values of the negative-sign ellipse correspond to the ones of the positive-sign ellipse and the degenerate solution has no effect.\footnote{Note that again the opposite side of the ellipse with respect to $\stheta$ has to be included because of the correlation between $\deltacp$ and $\stheta$.} In other cases, only a fraction of new values is moved into the error bars of the original solution, which means that the CP coverage is in-between the dashed curves (in the upper row of \figu{hkcpprec}) and their doubles. This complex interaction of the original and opposite-sign solution leads to the irregular CP pattern after inclusion of the degeneracy for $\stheta=0.01$.
\end{description}

In summary, the CP patterns for \JHFHK\ are much more complicated than the ones for the first-generation superbeams. However, resolving the $\mathrm{sgn}(\ldm)$-degeneracy would in this case help the CP performance very much and could even double the precision for intermediate values of $\stheta$.

\section{Summary and conclusions}
\label{sec:summary}

In this study, we have investigated the mass hierarchy, CP violation, and CP precision measurements at neutrino superbeams as function of the true value of $\deltacp$. We have used a likelihood analysis including systematics, correlations, degeneracies, spectral information, and a realistic detector simulation. Since the results from such an analysis are very difficult to interpret, we have chosen bi-rate graphs to illustrate many CP-phase dependent properties. Bi-rate graphs are based upon bi-probability graphs, but they use the total charged-current appearance rates instead of the oscillation probabilities in the bi-probability graphs. Since error bars can be attached to each point in a bi-rate graph, one can directly read off quantitative estimates for many effects. Furthermore, they are especially useful for off-axis superbeams, because these are ``quasi-monochromatic'' narrow band beams, which only carry little spectral information. Therefore, to a first approximation, only one ellipse in the bi-rate picture corresponds to one choice of energy and baseline. By comparing the bi-rate graphs with the complete statistical simulation, we have demonstrated that the role of the true value of $\deltacp$ in the mass hierarchy, CP violation, and CP precision measurements at superbeams can be understood in terms of these graphs. Though it can be used as a powerful tool in many cases, the bi-rate picture does not always exactly fit the complete calculations, since in some cases there are significant contributions of some spectral information, the disappearance channels, systematics, and correlations other than between $\stheta$ and $\deltacp$. We have shown these limitations at the example of the systematics domination of \JHFHK\ for small values of $\stheta$.

For the {\em mass hierarchy determination}, we have demonstrated that the true value of $\deltacp$ affects the potential to measure the mass hierarchy in a way that the $\stheta$-sensitivity reach is worst close to $\deltacp=\pi/2$ (conservative reach) and best close to $\deltacp=3 \pi / 2$.  Furthermore, the best case $\stheta$-sensitivity reach (close to  $\deltacp=3 \pi / 2$) hardly depends on the true value of $\sdm$, whereas the conservative case $\stheta$-sensitivity reach (close to  $\deltacp=\pi/2$) is highly affected by large values of $\sdm$. Since the true value of $\deltacp$ is given by Nature, it is often convenient to use the conservative reach for $\deltacp=\pi/2$ in order to determine what an experiment can do in either case. Eventually, the sensitivity to the normal mass hierarchy is better than the one for the inverted mass hierarchy. All of these effects can be qualitatively demonstrated with the help of bi-rate graphs.

The {\em sensitivity to CP violation} depends on the relative positions of the original and degenerate CP conserving values ($0$ and $\pi$) compared with the considered CP violating value. If the error bars at a CP violating solution in the bi-rate graph overlap one of the CP conserving solutions, then the sensitivity to CP violation will be destroyed. The asymmetry of the CP violation sensitivity between the first/second  $\deltacp \in  \, ]0, \pi[$ and third/fourth quadrants $\deltacp \in \, ]\pi, 2\pi[$ is caused by the $\mathrm{sgn}(\ldm)$-degeneracy. For the normal mass hierarchy, primarily the sensitivities to CP violation in $\deltacp \in  \, ]0, \pi[$ are affected by the $\mathrm{sgn}(\ldm)$-degeneracy, and for the inverted mass hierarchy, primarily the sensitivities to CP violation in $\deltacp \in \, ]\pi, 2\pi[$.  Note that CP violation measurements are very difficult for first-generation superbeams. In fact, only the combination of \JHFSK\ (neutrino running only) combined with \NUMI\ (antineutrino running only) would only have a sensitivity close to maximal CP violation $\deltacp \simeq \pi/2$ (for inverted mass hierarchy) or $\deltacp \simeq 3 \pi/2$ (for normal mass hierarchy).

Unlike the CP violation measurement, the {\em CP precision measurement} indicates if one could learn something about $\deltacp$ at all. 
For example, one may not be able to detect CP violation, since $\deltacp$ is too close to $0$ or $\pi$, but one may still be able to exclude certain ranges for $\deltacp$. We have used the ``CP coverage'' as a useful quantity to describe the $\deltacp$-performance of an experiment. It is the range of $\deltacp$-values (as fraction of the CP circle) which fit a chosen simulated value (true value) of $\deltacp$. Therefore, the concept of the CP coverage takes into account the cyclicity of the $\deltacp$-dependence. In addition, it can describe $\deltacp$-exclusion measurements, as well as high-precision measurements.
 The CP coverage strongly depends on the true value of $\deltacp$ itself, and this dependency results in rather complicated ``CP patterns''. The CP patterns depend on $\stheta$ and the presence of the $\mathrm{sgn}(\ldm)$-degeneracy, and can be qualitatively understood in terms of bi-rate graphs. 
The strong dependence on the true values of $\deltacp$ itself implies that one has to be careful when quoting or comparing CP precision performances of different experiments. For example, an experiment, such as a first-generation superbeam, may not supply any new information on $\deltacp$ in a wide range of true values of $\deltacp$. Thus, it is always useful to show and compare the complete CP patterns for different experiments. 
In particular, we have also demonstrated that for the normal mass hierarchy, only the precisions for $\deltacp \in  \, ]0, \pi[$ are affected by the $\mathrm{sgn}(\ldm)$-degeneracy, and for the inverted mass hierarchy, only the precisions for $\deltacp \in \, ]\pi, 2\pi[$ for the first-generation superbeams. Thus, the $\mathrm{sgn}(\ldm)$-degeneracy causes an  asymmetry between the first/second  $\deltacp \in  \, ]0, \pi[$ and third/fourth quadrants $\deltacp \in \, ]\pi, 2\pi[$. For the superbeam upgrades, the situation is more complicated, since the error bars become very small and the $\mathrm{sgn}(\ldm)$-degeneracy may or may not be present at the chosen confidence level. Especially for intermediate values of $\stheta$, the presence of the $\mathrm{sgn}(\ldm)$-degeneracy causes very irregular CP patterns. In addition, especially for high precision instruments, the appearance of all degeneracies as functions of the confidence level is the reason that the CP precision over-proportionally depends on the chosen confidence level. Thus, it is important that the same confidence level be used for a comparison of different experiments.

Finally, bi-rate graphs have been demonstrated to be a useful tool, which links the qualitative predictions of the bi-probability picture with complete quantitative evaluations including spectral information, systematics, correlations, and degeneracies. For example, the complicated CP patterns of the CP precision measurements can be understood in terms of bi-rate graphs. Of course, it would also be useful to extend this discussion to neutrino factories. However, neutrino factories have a lot more spectral information than superbeams, and therefore, their treatment in terms of bi-rate diagrams will be much more complicated. Nevertheless, one can still find many of the  qualitative features observed in this work in neutrino factory measurements.

\subsection*{Acknowledgments}

I would like to thank Steve Geer, Deborah Harris, Patrick Huber, Manfred Lindner, and Tommy Ohlsson for useful discussions. In addition, I especially would like to thank Stephen Parke for extensive discussions and valuable comments, and the Fermilab theory group for their warm hospitality during my stay within the summer visitor program.

This work has been supported by the ``Sonderforschungsbereich 375 f\"ur Astro-Teilchenphysik der Deutschen Forschungsgemeinschaft'' (SFB 375 of DFG), the ``Studienstiftung des deutschen Volkes'' (German National Academic Foundation), and the Fermilab summer visitor program.

\end{document}